\documentclass[conference]{IEEEtran}
\usepackage{fullpage}
\usepackage{url,times}
\usepackage{graphics,graphicx}
\usepackage{subfigure}
\usepackage{amssymb,amsmath,amsbsy, amstext}
\usepackage{color}

\begin{document}

\title{Aligned Virtual Coordinates for Greedy Routing in WSNs}

\author{Ke Liu and Nael Abu-Ghazaleh \\
CS Dept., SUNY Binghamton \\
\url{{kliu,nael}@cs.binghamton.edu}}

\maketitle

\begin{abstract}
Geographic routing provides relatively good performance at a
much lower overhead than conventional routing protocols such as 
AODV.  However, the performance of these protocols is impacted by
physical voids, and localization errors.  Accordingly, virtual
coordinate systems (VCS) were proposed as an alternative approach that
is resilient to localization errors and that naturally routes around
physical voids.  However, we show that VCS is vulnerable to different
forms of the void problem and the performance of greedy routing on VCS
is worse than that of geographic forwarding.  We show that these
anomalies are due to the integral nature of VCS, which causes
quantization noise in the estimate of connectivity and node location.
We propose an aligned virtual coordinate system (AVCS) on which the
greedy routing success can be significantly improved.  With our
approach, and for the first time, we show that greedy routing on VCS
out-performs that on physical coordinate systems even in the absence
of localization errors. We compare AVCS against some of the most
popular geographical routing protocols both on physical coordinate
system and the virtual coordinate systems and show that AVCS
significantly improves performance over the best known solutions.


\end{abstract}

\IEEEpeerreviewmaketitle

\section{Introduction}

\label{sec:introduction} 

In contrast to traditional ad hocrouting  protocols such as
AODV~\cite{paper:aodv}, Geographical routing
~\cite{paper:gfg,paper:gpsr,paper:gcrp,paper:practical_gr},
provides attractive properties for WSNs.  It operates via local
interactions among neighboring nodes and requires little state
information that does not grow with the number of communicating nodes.
In these algorithms, nodes exchange location information with their
neighbors.  Packets addressed to a destination must provide its
location.  At every intermediate hop, the subset of the neighbors that
are closer to the destination is called the forwarding set (FS).
Routing simply forwards a packet a node in FS.  This
process is repeated greedily until the packet reaches the destination.
Thus, interactions are localized to location exchange with direct
neighbors and there is no need for global identifiers.

Geographical routing protocols suffer from significant problems under
realistic operation.  First, voids --intermediate nodes whose FS
relative to a destination is empty-- can cause the greedy algorithm to
fail \cite{paper:gfg,paper:gpsr,paper:bphole,paper:gcrp}. Voids
require a somewhat complex and inefficient complementary routing
algorithm (e.g., perimeter routing) that is invoked when they are
encountered, which requires more information in addition to the
location of neighbors \cite{paper:practical_gr}. Moreover, geographic
routing has been shown to be sensitive to localization
errors~\cite{paper:errorgf}, especially in the perimeter routing
phase~\cite{paper:practical_gr,paper:errorfr}; such errors can cause
routing anomalies ranging from suboptimal paths to loops and failure
to deliver packets. Making geographical routing protocols practical is
extremely difficult~\cite{paper:practical_gr}.

Routing based on Virtual Coordinate Systems (VCS) has been recently
proposed~\cite{paper:vcembed,paper:lcr,paper:vcsim,paper:vcap,paper:bvr}
to address some of the shortcomings of geographic routing.  A VCS
overlays virtual coordinates on the nodes in the network based on
their distance (typically in terms of number of hops) from fixed
reference points; the coordinates are computed via an initialization
phase.  VCS coordinates serve in place of the geographic location for
purposes of geographic forwarding; that is, in these algorithms the FS
is the set of nodes that are closer to the destination than the
current node, based on a function that computes distance between
points in coordinate space (e.g., Cartesian distance, or Manhattan
distance).  Because it does not require precise location information,
VCS is not sensitive to localization errors.
Further, it is argued that VCS is not susceptible to conventional
voids because the coordinates are based on connectivity and not
physical distance \cite{paper:lcr}.  On the negative side, VCS may
be sensitive to collisions and or signal fading effects in the
initialization phase.  Furthermore, the initialization phase requires
a flood from each reference point.  Finally, the coordinates should
be refreshed periodically if the network is dynamic.  These issues are
not present in geographical routing.  We call both geographic and
virtual coordinate routing {\em geometric routing}.

Most existing research work in geometric routing protocols
concentrates on optimizing the complementary routing algorithm such as
perimeter routing\cite{paper:gpsr,paper:gfg,paper:practical_gr}, or
backtracking in VCS
\cite{paper:vcembed,paper:vcsim,paper:lcr,paper:bvr,techrpt:hgr}.  The
dominant part of the geographical routing, greedy forwarding, is
largely ignored.  However, since performance during greedy forwarding
phase is much better than during the complementary phase, it stands to
reason that increasing the percentage of paths that can be routed in
the greedy mode improve overall performance.  This is the primary
advantage of our approach.  To provide the necessary context to
describe our contribution, Section~\ref{sec:related} reviews related
work.

The first contribution of the paper is to identify, for the first
time, the {\em VCS forwarding void} that arises due to the {\em
  virtual coordinate quantization noise}.  More specifically, since
virtual coordinates are based on integer number of hops to the
reference nodes, they represent a coarse approximation of node
location; several nodes which are not close to each other may share
the same coordinates, or have the same distance to a destination.
This leads to a special type of forwarding voids which cause greedy
forwarding to fail even without the presence of physical voids.  We
explain how this problem arises and analyze the frequency of
occurrence.  Other routing
anomalies that arise in VCS systems~\cite{techrpt:hgr}, are also
contributed to by the quantization noise.  These problems are
discussed in Section~\ref{sec:vcsproblems}.

The second contribution of this paper is an aligned virtual coordinate
system (aligned VCS, or AVCS) on which the greedy forwarding phase
becomes more immune to quantization noise.
Specifically, nodes align themselves to a non-integral coordinate
point that is a function of not only their own coordinates, but also
of those of their neighbors.  This alignment process significantly
reduces the quantization noise, and alleviates many of the VCS
forwarding voids, especially under uniform node density.  AVCS is
presented in Section~\ref{sec:avcs}.

We use simulation to compare the performance of greedy routing
protocols on different coordinate systems, such as geographic
coordinate system (GeoCS), VCS and the aligned VCS in
Section~\ref{sec:experiment}.  The aligned VCS is able to deliver
packets in the greedy forwarding mode at a much higher rate than the
other coordinate systems we analyze, while achieving a good path
quality that approaches that of stateful Shortest Path (SP) routing
such as AODV.  Using VCS, the percentage of time that the expensive
complimentary void traversal algorithm is invoked is significantly
reduced.  Please note that aligned VCS optimizes the greedy phase, and
can inter-operate with any complimentary algorithm for packets that
reach voids.  The path stretch of greedy forwarding is shown to be as
good as SP experimentally.  We present some concluding remarks and
future work in Section~\ref{sec:conclusion}.

\section{Background and Related Work}
\label{sec:related}

Stateful hop-count based routing protocols such as AODV \cite{paper:aodv}, are
commonly-used in Ad hoc networks.  A variant, called Shortest Path (SP), can be used in sensor networks: in
SP, data sinks send periodic network-wide beacons
(typically using flooding). As nodes receive the beacon, they set their next
hop to be the node from which they received the beacon
with the shortest number of hops to the sink. Thus, with a single
network wide broadcast, all nodes can construct routes to the
originating node. This functionality is convenient for data gathering
applications where there is a single data destination. SP can provide
the optimal path in terms of path length.  However, it is a stateful
and reactive protocol: for each data sink, the forwarding path is
needed before data transmission can begin.  The storage it requires
increases with the number of sinks in the network. Furthermore, it is
vulnerable to mobility or other changes in the topology.

To counter these disadvantages, stateless geometric
routing protocols were proposed.  GFG~\cite{paper:gfg}, and the very
similar GPSR~\cite{paper:gpsr}, are the earliest and most widely used
of this class of protocols.  They consist of a Greedy
Forwarding (GF) phase where each node forwards packets to the neighbor
that will bring the packet closest to the destination.  
Each node tracks only the location information of its
neighbors.  Based on this information, for a packet with a given
destination, a node can determine the set of neighbors 
closer to the destination than itself; this set is
called the {\em forwarding set} for this destination.  GF proceeds by
picking a node from this set, typically the closest to the
destination.

It is possible that GF fails, if the forwarding set is empty: a {\em
  void} is encountered.  A complementary phase of the algorithm is
then invoked to traverse the void.  Typically, face routing or
perimeter routing; this is an approach based planar graph theory.
The general idea is to attempt to route around the void using a right
hand rule that selects nodes around the perimeter of the void (details
may be found in the original paper~\cite{paper:gpsr}).
This approach is continued until a node closer to the destination than
the void origin is encountered; at this stage, operation switches back
to greedy forwarding.  However, a problem arises if the perimeter
routing intersects itself -- there is a danger that the packet gets
stuck in a loop.  Thus, a technique for planarizing the graph to avoid
the use of intersecting edges is needed: {\em Relative Neighborhood
  Graph (RNG)} and {\em Gabriel Graph (GG)} are 2 kinds of such
planarization techniques.

GPSR and other geographic routing protocols are vulnerable to
localization errors.  The localization process has built in tolerances
and, in general, location information is not precise. The degree of
error in the location estimate depends on the localization mechanism
(an error up to 40\% of the radio range is considered a common
case). Since GPS devices are costly, they may not be feasible for
sensor networks; often, localization algorithms are employed that
significantly increase the uncertainty in the location estimate (e.g.,
~\cite{paper:aps,paper:hightower01,paper:robustloc-mobicom2002}).
Both the greedy forwarding and face routing phases are susceptible to
localization errors~~\cite{paper:errorgf,paper:errorfr}. While some
approaches to tolerate location errors have been suggested, in
general, this remains a weakness of this class of protocols. Further,
the paths constructed by face routing are typically not the best path
available to cross the void; they can be extremely inefficient,
especially if the network is dense.  Thus, additional routing
protocols have attempted to optimize the face routing phase of
operation~\cite{paper:gcrp,paper:bphole,paper:glider}. However, most
of these works optimize face routing in term of path quality, but tend
to increase the overhead and the complexity.  They do not address the
effect of location errors on the improved schemes. 

Routing based on a coordinate system, rather than location, was first
proposed by Rao et al~\cite{paper:grnoloc}.  However, this approach
requires a large number of nodes to serve as virtual coordinate anchor
nodes (sufficient to form a bounding polygon around the remaining
sensors).  The drawback of having many reference points is that
forming coordinates requires a long time to converge and a very high
network density; the same is true for the overhead to refresh
coordinates.  Instead of using the virtual coordinates directly for
routing, they use them to estimate location for use in geographic
routing.  Reachability is expected to be an issue in this protocol as
geographic location is approximate; recall that it has been shown that
both the greedy forwarding and the face routing phases of geographic
routing are susceptible to localization errors.  Similar approaches
that use VCS to aid localization have been also used by other
works~\cite{paper:gc,paper:aps}.  Note that these works collapse the
original VCS coordinates back into 2 geographic coordinates for the
purpose of routing.  

GEM~\cite{paper:gem} proposed the routing based on a virtual
coordinate system. A virtual polar coordinate space (VPCS) is used for
localizing each node in network. A tree-style overlay is then used for
routing. Thus, GEM is not stateless.  Further, using the tree overlay
results in poor path quality.  Since it uses the VPCS to localize the
network first, it tolerates only up to $10\%$ localization
error~\cite{paper:gem}.

Caruso et al recently proposed the Virtual Coordinate assignment
protocol (VCap) \cite{paper:vcap}.  Several similar protocols are also
proposed\cite{paper:vcembed,paper:vcsim,paper:lcr,paper:bvr,techrpt:hgr}. In
this approach, coordinates are constructed in an initialization phase
relative to a number of reference points. Following this
initialization phase, packets can be routed using the Greedy
Forwarding principles, replacing node location with its coordinates:
the forwarding set consists of neighbors whose coordinates are closer
(different distance functions have been proposed) to the destination
than the current node. The paper advocates the use of 3 reference
points to assign the virtual coordinates, constructing a 3-dimensional
VCS. We showed that this 3D VCS may not sufficient to map the network
effectively\cite{techrpt:hgr}. VCap, even with 4 coordinates, is worse
than GPSR both in delivery ratio (node pair reach-ability) and path
quality. We also demonstrate and experimentally show that Greedy
Forwarding on 3D VCap is significantly worse than normal
geographically based Greedy Forwarding.

Qing et al proposed a similar protocol, called Logical Coordinate
Routing (LCR), to VCap with 4 reference nodes (4D) each located at a
corner for a rectangular area~\cite{paper:lcr}. The choice of the
number of reference nodes was not explicitly explained; however, we
note that 4 corner nodes are sufficient to form a bounding polygon of
a rectangular area. The authors suggested a backtracking approach to
deliver packet when facing any routing anomalies.  This approach
requires that each hop in the forwarding path of each packet to be
recorded. LCR and other VCS algorithms can benefit from the proposed
Aligned VCS idea to improve the performance of the greedy phase.

The use of a Manhattan-style distance was proposed by Rodrigo et al in
BVR\cite{paper:bvr}. On a VCS with much more reference nodes
(typically 10 to 80), BVR suggested a different backtracking approach
to forward packets back to the reference node closest to the
destination when greedy forwarding fails. As we show in this paper,
neither Manhattan distance nor the one proposed in BVR\cite{paper:bvr}
(which we call semi-Manhattan distance) are necessarily better
measures of than Euclidean distance. The use of a high number of
reference nodes requires proportionately higher overhead in terms of
communication and state, both during set up and refresh of the coordinates.



Papadimitriou and Ratajczak~\cite{paper:gfconjecture} conjecture that
every planar 3-connected graph can be embedded on the plane so that
greedy routing works.  If this conjecture holds, a coordinate system
where a guaranteed greedy routing may exist for any connected network.
Our work may be considered a step towards this goal.

\section{Greedy Forwarding in VCS}
\label{sec:vcsproblems}
The Virtual Coordinate System (VCS) for wireless routing was
introduced by VCap~\cite{paper:vcap}. It is attractive for use in
environments where the advantages of geographic routing are desired,
but are not possible due to localization errors.  The virtual
coordinates of each node in the network are set up by tracking the
number of hops from several virtual coordinate anchors.
A network using SP with $N$ sinks can be considered an $N$-dimensional
VCS as the distance to each of the sinks is tracked.  The
authors~\cite{paper:vcap} argue that for a 2 dimensional geographical
coordinate system (GeoCS), a 3-dimensional VCS is sufficient to
accomplish effective Greedy Forwarding (GF). In \cite{techrpt:hgr} we
showed that in practice this is not possible, and low delivery ratio
is achieved, unless at least 4 dimensional VCS is used.

Although VCS appears to overcome voids because it is based on
connectivity rather than geographical location; it does not achieve
perfect greedy routing.  Far from it, its greedy routing phase fails
more often than geographical routing in most situations.  Several
routing problems that arise with VCS that result from voids perturbing
the coordinate space\cite{techrpt:hgr}.  In this paper, we show
another problem in greedy forwarding in VCS that arises even when
there are no physical voids.  In all these problems, the result is
that a packet reaches a node with no neighbors closer to the
destination than itself; a local minimum is reached and greedy
forwarding fails.

\begin{figure}[t]
\includegraphics[width=0.48\textwidth]{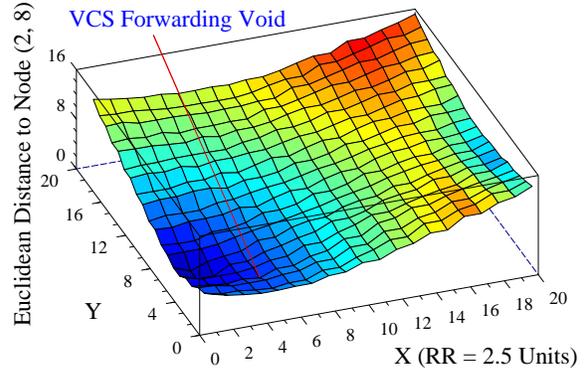}
\caption{Euclidean Distance Map: Forwarding Void in 4D VCS}
\label{fig:fv1}
\end{figure}

Consider a 4D VCS, set up according to the VCAP scheme
\cite{paper:vcap,paper:lcr,techrpt:hgr}.  Further, consider a set of nodes
A, B and C with virtual coordinates as V(A), V(B) and V(C), where A
and B are neighbors, and B and C are also neighbors. According to the
design of VCS, we have
\begin{eqnarray*}
0\leq&abs(V(A)_i - V(B)_i)&\leq1\\
0\leq&abs(V(B)_i - V(C)_i)&\leq1
\end{eqnarray*}
If we also have
$$Distance(A, B) = Distance(A, C)$$ where $Distance(Node 1, Node 2)$
can be measured in different ways such as Euclidean distance
\cite{paper:vcap,paper:lcr} and Manhattan-style distance
\cite{paper:bvr}, then a packet from node C may not be delivered to
node A even if there is a path through node B.  In simulations, we
observed such conditions arising often, and in many scenarios are the
primary cause of undeliverable packets.  For example, in a simulation
with 400 nodes that are uniformly deployed with an average density of 10 neighbors per node,  greedy forwarding between nearly 20\% of the pairs of the nodes failed, with 4D-VCS.  For 3-D VCS, roughly 40\% failed.
For example, the virtual
coordinates of A, B and C are
\begin{eqnarray*}
V(A)&=[3, 9, 7, 11]\\
V(B)&=[2, 9, 8, 11]\\
V(C)&=[3, 8, 8, 11]
\end{eqnarray*}
So the $Distance(A, B)=\sqrt{2}$ and $Distance(A, C)=\sqrt{2}$
measured in Euclidean Distance, while the $Distance(A, B)=2$, and $Distance(A,
C)=2$ measured in Manhattan distance. Even though the virtual coordinates
of A, B and C satisfy the design of VCS and they are in the same
neighboring chain, greedy forwarding fails at node C. We
demonstrate this problem as a distance map in figure \ref{fig:fv1},
where the X and Y denotes the physical location of each node,
and Z denotes the Euclidean distance
of each node of virtual coordinates to the node locates at (2, 8)
(node A).
As we can see, the node located at (5, 4) (node C) has no neighbor
closer to the destination (node A) per either the Euclidean or
Manhattan distance of virtual coordinates.  We call this problem a
{\em VCS forwarding void}. A more serious forwarding void may be found
in the region around location (19,6) where the virtual coordinates of
all nodes around it are further away away from the destination in this
figure.

The reason behind the VCS forwarding void is the {\em virtual
coordinate quantization noise}.  Two nodes may receive the same virtual
coordinate value $x$ at a given dimension, their physical distances
$Dis$ to the anchor node providing the dimension beacon may match
\begin{equation*}
\left\{ \begin{array}{l l} 0< Dis \leq RR & \text{~for~} x=1 \\
\frac{1}{2}x\times RR < Dis \leq x\times RR &\text{~for~} x\geq 2
\end{array}\right.
\end{equation*}
where the $RR$ denotes radio range.  For each hop (1 in $x$ value), the
noise would be at most $\frac{1}{2}RR$ under unit disc assumptions. As
the value of $x$ goes higher, the noise becomes bigger. Since the virtual
coordinate value just reflects the distance of a node to some anchor
measured in number of hops, the noise affects the greedy forwarding
significantly in the networks with a large number of nodes.

\begin{figure}[t]
\includegraphics[width=0.48\textwidth]{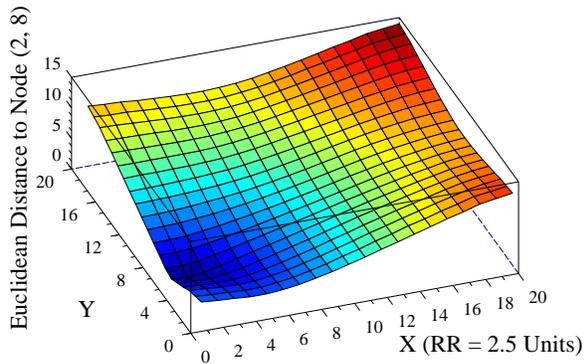}
\caption{Distance Map of Aligned VCS Example} \label{fig:afv1}
\end{figure}

The quantization noise comes from the hop-count nature of the VCS
which uses integer values to approximate continuous physical
locations.  A more accurate approach to mapping the network into a VCS
is needed. The virtual coordinate value should reflect not only the
distance to the anchor nodes, but also the connectivity of the
neighborhood. We propose such an approach which we call aligned VCS
(described in Section~\ref{sec:avcs}). The use of aligned VCS on the
20x20 grid network causes the distance map as figure \ref{fig:afv1}
where the distances of all nodes to the destination node is
continuously decreasing, allowing more effective greedy forwarding.

\section{The Design of Aligned VCS}
\label{sec:avcs}


For geographical routing protocols, a relative high density can
resolve many routing anomalies such as physical voids (ignoring the
effect of localization errors). In contrast, high density may not
resolve the routing anomalies in VCS \cite{techrpt:hgr}; it may even
exacerbate them due to the quantization noise problems such as {\em
VCS forwarding void} presented in section \ref{sec:vcsproblems}.  The
complementary solutions for a void avoidance are not effective as
described in section \ref{sec:related}. Consequently, the performance
of geographical routing protocols on VCS is not as good as it on GeoCS
with precise location information or any stateful routing protocols.
A successful geographical routing protocol should be
stateless, tolerate localization errors, and use greedy forwarding as
much as possible in order to perform as well as on GeoCS or stateful
routing protocols.

The integral-valued nature of the VCS contains quantization noise.
Further, the discrete nature of the integral coordinates makes it
easier to reach routing anomalies where all neighbors are of equal
coordinates or equal distance to the destination.  The goal of the
proposed Aligned Virtual Coordinate System (AVCS) is to provide more
representative and continuous coordinate and connectivity information.

\subsection{Aligned VCS}

The virtual coordinates of each node in the VCS is set up as the
vector of the hop-counts from several anchor nodes.  For any node A
with a virtual coordinate vector $V(A)$, the virtual coordinate vector
$V(N)$ of any neighbor N of it would satisfy
\begin{equation*}
0\leq abs(V(N)_i - V(A)_i) \leq 1 \text{~for~all~} i
\end{equation*}
The integral value of virtual coordinate at dimension $i$ cannot
discriminate between two nodes with the same coordinate value. The
value can only tell us what level of a routing trees rooted at the
anchor nodes the two nodes belong to.  However, by considering the
neighbor information, which is generally different for the two nodes,
a more effective and discriminating coordinate value can be achieved.
More specifically, alignment refers to the process of computing the
coordinates of a given node as a function of its own coordinates and
the coordinates of the neighboring nodes.  Thus, given the same
initial coordinate value, a node A with {\em neighbors that average a
smaller virtual coordinate value} than another node B, is closer to
the root of the routing tree than B is.  In this case, B has {\em
neighbors which average bigger virtual coordinate value} is aligned
further away from the root.  Similarly, a node with {\em neighbors
which average roughly the same virtual coordinate value} is aligned
towards the middle.  A possible alignment function that we consider
produces an aligned virtual coordinate vector $AV(A)$ of a node $A$ as
follows.
\begin{equation}
AV(A)_i^d = \frac{\frac{\sum_{j=1}^n{AV(N_j)_i^{d-1}}}{n} +
  AV(A)_i^{d-1}}{2}
\label{eq:self_factor}
\end{equation}
where the $i$ is the $i^{th}$ virtual dimension, the $N_j$ is the
$jth$ neighbor of A, and $n$ is the number of neighbors of $A$.  $d$
is the depth of aligned virtual coordinates.  The aligned virtual
coordinates with depth 0 stand for the original integral hop-counter
virtual coordinates value.  A depth of 1 indicates averaging
coordinates among one-hop neighbors, and a depth of $n$ reflects
taking into account neighbors that are $n$ hops away from the node.
An alternative equation for aligned virtual coordinates might be
\begin{equation}
AV(A)_i = \frac{\sum_{j=1}^n{AV(N_j)_i^{d-1}}+AV(A)_i^{d-1}}{n+1}
\label{eq:average_factor}
\end{equation}
The estimates of the two functions are not likely to be significantly
different.

\subsection{Distance Measurement on Aligned VCS}

In geographical routing, each data packet has to carry the coordinates
value of the destination node. To make a routing decision in greedy
forwarding, the distance from current node to the destination is
compared to those of one-hop neighbors.  The closest neighbor to the
destination would be chosen as the next hop to which the packet would be
forwarded. This process repeats until the packet arrives at the
destination.

  For a network where nodes do not move frequently, the stability of
nodes keeps the aligned VCS stable. So for routing decisions, the
updated aligned virtual coordinates of destination is not difficult to
obtain. But if nodes with high mobility are common, the aligned
virtual coordinates of nodes would turn stale quickly. For this
consideration, the aligned virtual coordinates of any node as
destination is not used for routing, which means that each data packet
carries only the integer value virtual coordinate of the destination
($V(dst)$ or say, $AV(dst)^0$). The aligned virtual coordinates are
used only for distance measurement locally.  So the Euclidean distance
from a given node $X$ to the destination $dst$ is measured as
\begin{equation}
EDis(X, dst)^d = \sqrt{\sum_{i=1}^n{(AV(X)_i^d - V(dst)_i)^2}}
\label{eq:edis}
\end{equation}
For a 4D VCS, $n=4$. And the Manhattan distance can be measured by
\begin{equation}
MDis(X, dst)^d = \sum_{i=1}^n{abs(AV(X)_i^d - V(dst)_i)}
\label{eq:mdis}
\end{equation}
Other distance functions such as the semi-Manhattan function used by
BVR \cite{paper:bvr} can be measured in a similar way.

It is important to note that AVCS is quite different from works that
use virtual coordinates to localize for use in geographic routing.
Those works attempt to measure physical distance and not align based
on connectivity.  Furthermore, they compress the VCS dimensions, no
matter how many, back into two geographic dimensions X and Y,
losing a significant amount of information.

\section{Experiment}
\label{sec:experiment}

In this section, we present an experimental evaluation that
illustrates the existing geographical routing protocols on physical
coordinates (GeoCS), virtual coordinates (VCS) and aligned virtual
coordinates (AVCS) systems. The evaluation tracks metrics such the
greedy ratio (portion of paths that can be routed using greedy
forwarding only) and average path length (path stretch relative to
SP). The reason to choose the greedy ratio as an evaluation metric is
that it reflects both the overhead and the performance of a stateless
routing protocol such as GPSR; the lower the greedy ratio, the more
frequently we need to use the more expensive and less efficient
complimentary perimeter routing.

\subsection{Experimental Setup and Preliminaries}

To enable scalability, we use a custom simulator written for this
study; the simulator abstracts away the details of the channel and the
networking protocols. Since our work targets functionality in the
control plane (not the data plane), we believe that hiding the
modeling details of that level should be better. The results of the
simulator validate well with the NS-2 simulator~\cite{ns2}.  However,
NS-2 does not allow scaling the simulation size to the network sizes
we want to study.

We study the impact of physical voids on geographical routing with
different coordinate systems through a number of scenarios.  For every
scenario, the greedy ratio and path stretch are determined as the
average of these values for every pairwise permutation of the nodes in
the network; that is, one test is done for sending a packet from every
node to every other node. If the packet is delivered through greedy
forwarding, it counted towards the greedy ratio.  We use SP, which
finds the optimal routing in terms of number of hops, as the baseline
for measuring path stretch.  Note that SP is stateful and expensive especially if the network is dynamic or the number of destinations large.

We implemented GPSR (with both GG and RNG planarization)
\cite{paper:gpsr,paper:practical_gr}, Shortest Path (SP), Greedy
Forwarding on VCS \cite{paper:vcap}, LCR~\cite{paper:lcr} and
BVR~\cite{paper:bvr} on 4D VCS, to study their performance against
Aligned VCS (AVCS).  In fairness to BVR, the original specification
suggests 10 to 80 reference points; however, we are interested in
evaluating it against the other schemes with similar assumptions on
the number of beacons.  In addition, we believe that such a large
number of reference points is impractical in many settings.  The
aligned VCS uses Equation (\ref{eq:average_factor}). The simulation
results show the 2 equations (\ref{eq:self_factor},
\ref{eq:average_factor}) perform almost identically (less than 0.5\%
difference).  The distance between any 2 nodes is measured through a
Euclidean manner as equation (\ref{eq:edis}) except BVR.

\subsection{Greedy Forwarding vs Shortest Path}
\label{sec:gfvssp}

The first study shows the performance of geographical Greedy
Forwarding (GF) and Shortest Path (SP) routing in terms of path
length.  These results form part of the intuition for our work.  More
specifically, the greedy component of geographic and VCS routing is
the close to the optimal SP performance.  Thus, increasing the success
of the greedy phase leads to improving the effectiveness of geographic
and VCS routing.

Since voids cause non-greedy routing, in this study we use a grid
deployment of 2500 nodes in a 50x50 units area.  Each node is placed
at the center of one grid. The impact of the density is studied
through varying the transmission range.  The path lengths of SP GF on
GeoCS (greedy phase of GPSR), 4D VCS and 4D Aligned VCS with depth 1
(AVCS d1) are shown in Table \ref{tab:intuition}.  GF on GeoCS
performs the same as SP routing which means in this grid deployment
without any physical voids, the stateless routing protocols perform as
well as stateful one.  When the radio range becomes higher, the greedy
forwarding on VCS becomes worse than GF due to the increased
quantization noise and the VCS forwarding anomalies. Aligned VCS does
not suffer from this problem.

\begin{table}[h]
\begin{small}
\begin{tabular}{|c|c|c|c|c|}
\hline Neighbors \#& SP & GeoCS & 4D VCS & 4D AVCS (d 1)
\\\hline 3.92&1.0000&1.0000&1.0000&1.0000\\\hline
7.76&1.0000&1.0000&1.0000&1.0000\\\hline
11.60&1.0000&1.0000&1.0000&1.0000\\\hline
19.13&1.0000&1.0000&1.0494&1.0073\\\hline
26.57&1.0000&1.0000&1.0251&1.0010\\\hline
33.94&1.0000&1.0000&1.0409&1.0050\\\hline
44.84&1.0000&1.0000&1.0545&1.0035\\\hline
62.66&1.0000&1.0000&1.0850&1.0100\\\hline
73.17&1.0000&1.0000&1.0926&1.0074\\\hline
\end{tabular}
\end{small}

\caption{Path Stretch of Shortest Path (SP) and Greedy Forwarding
(on GeoCS, 4D VCS and 4D Aligned VCS)} \label{tab:intuition}
\end{table}

\begin{table}[h]
\centering
\begin{small}
\begin{tabular}{|c|c|c|c|}
\hline Neighbor \# & Perimeter Routing & BVR BT & LCR BT\\\hline
3.92&16.9200&2.1903&1.0000\\\hline
7.76&13.1800&2.0996&1.0000\\\hline
11.60&18.0365&2.1875&1.0000\\\hline
19.13&23.7324&2.1411&1.0525\\\hline
26.57&29.7087&2.1459&1.0277\\\hline
33.94&31.7657&2.1609&1.0480\\\hline
44.84&37.4030&2.1512&1.0564\\\hline
62.66&41.9031&2.1342&1.1198\\\hline
73.17&46.2990&2.1311&1.0982\\\hline
\end{tabular}
\end{small}

\caption{Path Stretch of Backtracking phase of geographical routing
protocols} \label{tab:backtracking}
\end{table}

Usually, the complementary solution for voids such as perimeter
routing and backtracking cause a significant path stretch relative to
SP. The path stretch of the complimentary algorithms in the previous
scenario were measured (Table~\ref{tab:backtracking}). The path
stretch of perimeter routing is extremely high but it is stateless,
requiring no more information than greedy forwarding. BVR is
considerably better, but still quite high, without requiring more
information. However, almost every time that backtracking was invoked
BVR required a scoped flood from the beacon which may cover half of
the network in a 4D VCS. Although the path stretch of LCR is the best
and approaches SP, it requires {\bf each} data packet to be recorded
by {\bf each} node in its forwarding path; this solution is
impractical for all but very lightly loaded networks.  From this
study, we can conclude that if the stateless routing can keep using
greedy forwarding, it can provide performance close to stateful
routing protocols while maintaining their desirable properties
(statelessness and low overhead).

\subsection{Effect of Voids}
\begin{figure}[t]
\centering
\includegraphics[width=0.33\textwidth,height=0.33\textwidth]{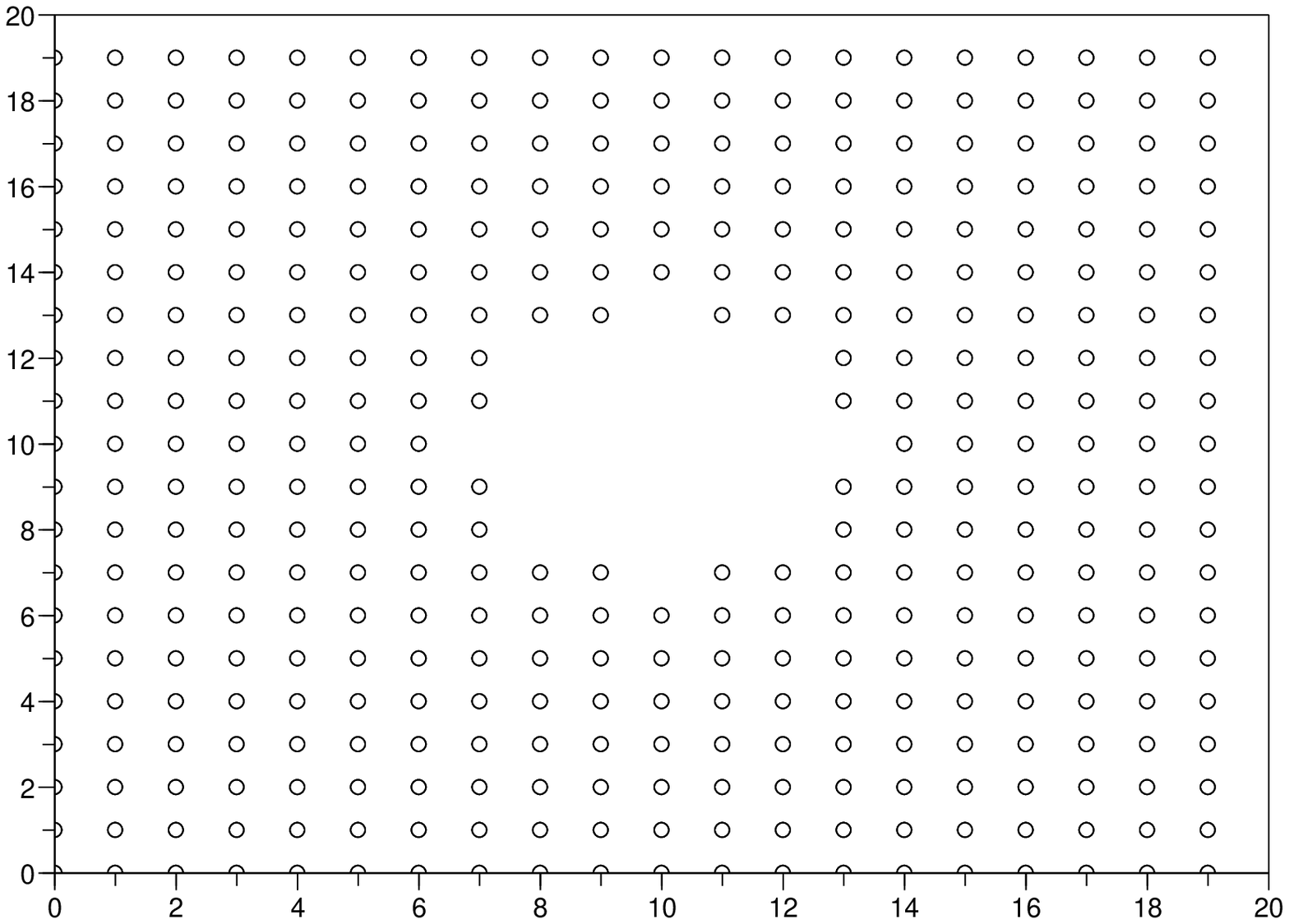}
\caption{Example of physical void: cycle hole 29/400}
\label{fig:hole29on400}
\end{figure}
First, we study the impact of the size of a single physical void on
performance.  We use a uniform
grid deployment of 400 nodes. The physical voids are created by taking
away some nodes in the center of area. An example of such a physical
void is shown in figure \ref{fig:hole29on400} where 29 nodes in the
center are taken away.  Later, we study random deployment scenarios.

\begin{figure}[t]
\includegraphics[width=0.45\textwidth]{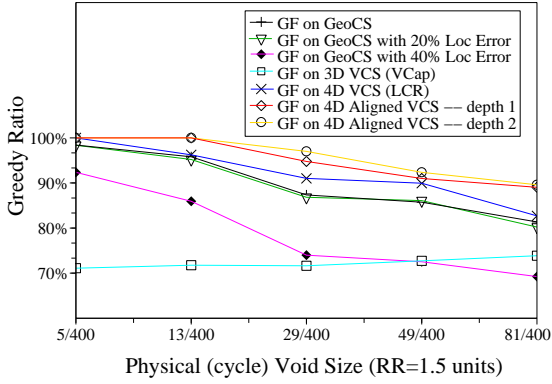}
\caption{Impact of void size: greedy ratio} \label{fig:hole_gr}
\end{figure}
\begin{figure}[t]
\includegraphics[width=0.45\textwidth]{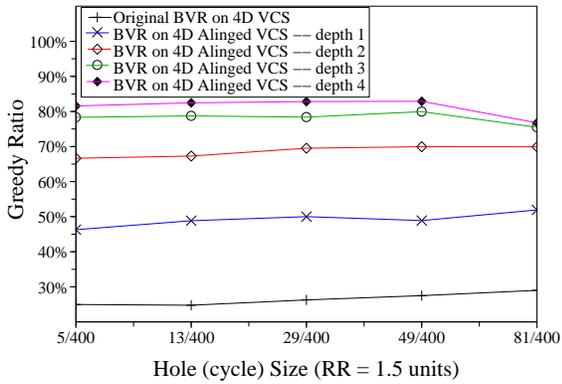}
\caption{Impact of void size on BVR: greedy ratio}
\label{fig:bvr_hole_gr}
\end{figure}
The greedy ratios of the geographical routing on different coordinate
systems are shown in Figure \ref{fig:hole_gr}.  GF on 4D VCS is able
to use greedy forwarding more often than GF on GeoCS. The greedy ratio
on 3D VCS is stable no matter what the size of the void is. The reason
may be that the anomalies of greedy forwarding on 3D VCS arise not
from voids primarily.  Although a localization error of around 20\% of
transmission range does not significantly affect the greedy ratio on
GeoCS, a 40\% localization error causes greedy ratio to drop
considerably which leads to higher path lengths and even routing
failure. Meanwhile, the geographical routing on VCS does not suffer
from localization errors. The greedy ratios of BVR on 4D VCS is shown
in figure \ref{fig:bvr_hole_gr}. As we can see, since the original BVR
is designed on a VCS with a relative higher dimensions (usually 10
$\sim$ 80), its greedy ratio with only four dimensions is very low.
However, increasing the number of coordinates higher, comes at a cost
of increased overhead in constructing the coordinates.  Even with
aligned VCS depth 3, its greedy ratio remains lower than the other
approaches.

\begin{figure}[t]
\includegraphics[width=0.45\textwidth]{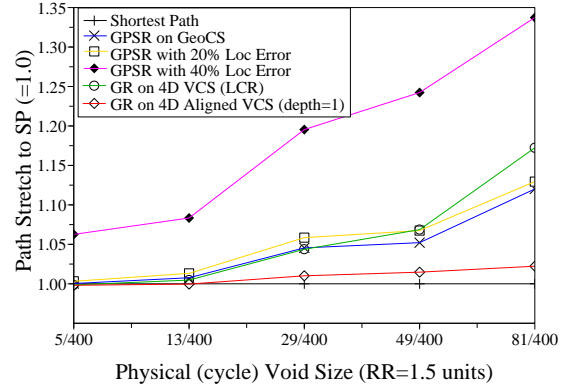}
\caption{Impact of void size: path stretch} \label{fig:hole_pl}
\end{figure}
\begin{figure}[t]
\includegraphics[width=0.45\textwidth]{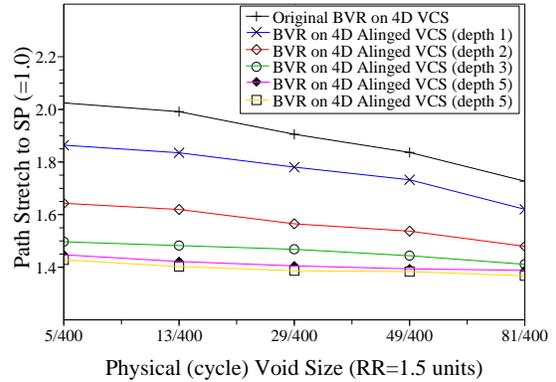}
\caption{Impact of void size on BVR: path stretch}
\label{fig:bvr_hole_pl}
\end{figure}
Figure \ref{fig:hole_pl} shows the path length of geographical routing
on different coordinate systems.  The primary observation from this
figure is that the greedy routing on VCS is better than that on GeoCS
even without any localization error, especially with a bigger physical
void. It can provide path stretch close to the optimal one even with a
large physical void. The path stretch of BVR is shown in figure
\ref{fig:bvr_hole_pl}. Similar to the study of greedy ratio, deeper
aligned VCS is, smaller path stretch is.

\begin{figure}[t]
\includegraphics[width=0.45\textwidth]{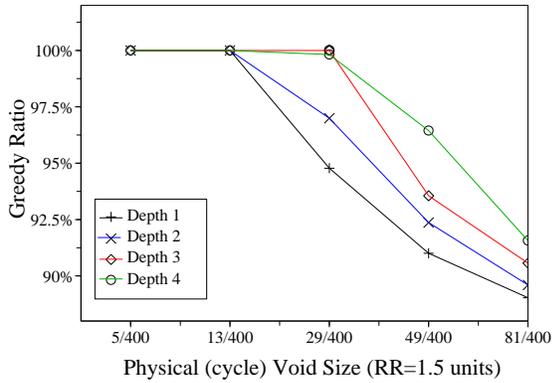}
\caption{Impact of void size on 4D Aligned VCS depth}
\label{fig:gr_depth}
\end{figure}
The impact of size of the void on AVCS for different
alignment depths is shown in Figure \ref{fig:gr_depth}. As expected,
the deeper the alignment, the higher the greedy ratio.  Although
deeper than 3 aligned VCS may still help, it may not be practical
if node mobility is present.

\begin{figure}[t]
\centering
\includegraphics[width=0.33\textwidth,height=0.30\textwidth]{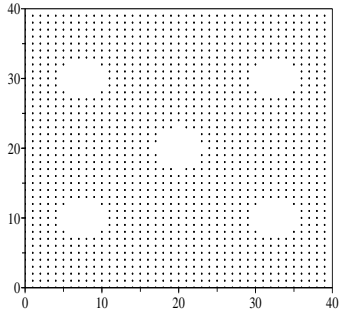}
\caption{Example of 5 physical holes in 40x40 grid area}
\label{fig:hole5}
\end{figure}

Next, we study the effect of multiple voids. 
In a 40x40 area
with grid deployment, nodes are removed to create
different number of holes. An example of 5 holes in such an area is
shown in Figure \ref{fig:hole5}.

\begin{figure}[t]
\includegraphics[width=0.45\textwidth]{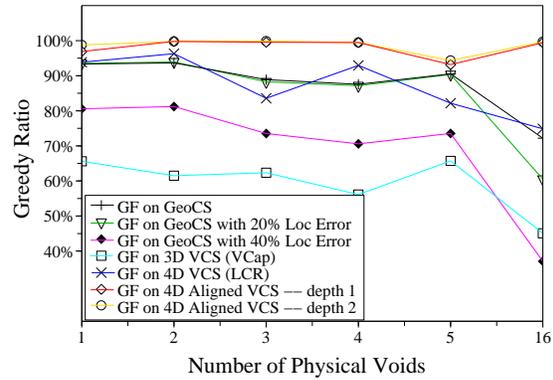}
\caption{Impact of multiple physical voids: greedy ratio}
\label{fig:mholes_gr}
\end{figure}

Figure \ref{fig:mholes_gr} shows the greedy ratio of geographical
routing on different coordinate systems as the number of physical
voids is increased. GF on 4D VCS has a similar performance as it does
on a GeoCS. The result does not support the use of VCS over GeoCS.
However, the greedy ratio using aligned VCS even with only depth 1
approaches 100\%, especially when the deployment of nodes is very
complicated (note the one with 16 physical holes).
\begin{figure}[t]
\includegraphics[width=0.45\textwidth]{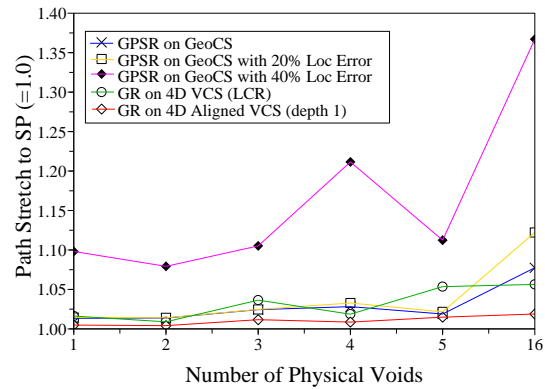}
\caption{Impact of multiple physical voids: path stretch}
\label{fig:mholes_pl}
\end{figure}
The study of path length in these scenarios is shown in Figure
\ref{fig:mholes_pl}.  The path stretch of greedy routing on 4D VCS
(either aligned or not) is just a little smaller than it on a GeoCS
for a small number of voids.  As the number of voids increases, the
greedy ratio of GeoCS drops, leading to a much higher path
stretch. Meanwhile, the greedy ratio of VCS is still high, and it
achieves good path quality as a result. With a 40\% localization
error, VCS significantly outperforms GeoCS.

\begin{figure}[t]
\includegraphics[width=0.45\textwidth]{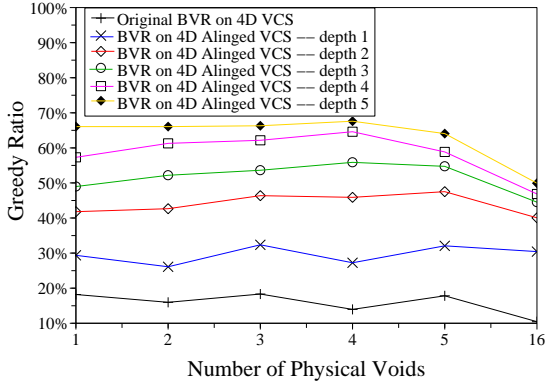}
\caption{Impact of multiple physical voids on BVR: greedy ratio}
\label{fig:bvr_mholes_gr}
\end{figure}

\begin{figure}[t]
\includegraphics[width=0.45\textwidth]{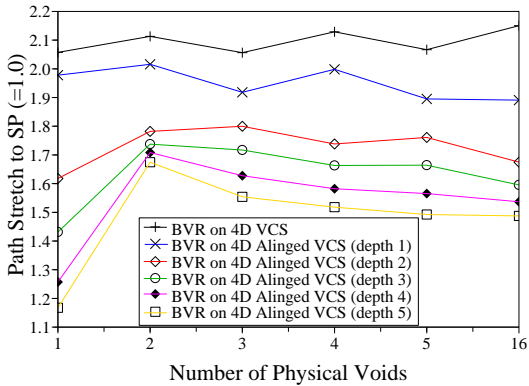}
\caption{Impact of multiple physical voids on BVR: path stretch}
\label{fig:bvr_mholes_pl}
\end{figure}

The performance of BVR in these scenarios is shown in Figure
\ref{fig:bvr_mholes_gr} and \ref{fig:bvr_mholes_pl}. Although the
aligned VCS greatly helps the performance of BVR , it is still worse
than other routing protocols which means its design requires a
large number of dimensions (anchor nodes). Because of this, we do not
include BVR in the remaining studies.

In unplanned wireless ad hoc and sensor networks deployments, the
placement of nodes is more likely to be random. We study scenarios
where 1600 nodes are deployed in a 30x30 area randomly.  The results
shown represent the mean of 20 randomly generated deployments, to
tightly bind the confidence interval. We study the impact of density
by increasing radio range.

\begin{figure}[t]
\includegraphics[width=0.45\textwidth]{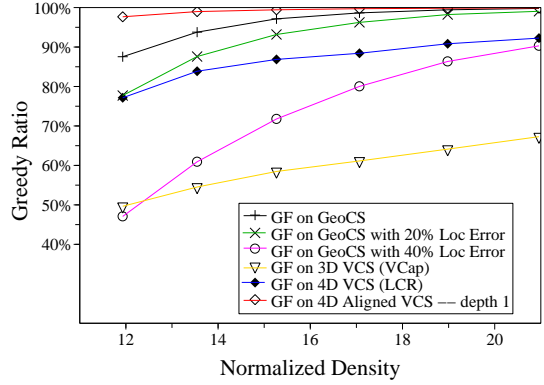}
\caption{Greedy Ratio of Reality: random deployment}
\label{fig:random_gr}
\end{figure}

\begin{figure}[t]
\includegraphics[width=0.45\textwidth]{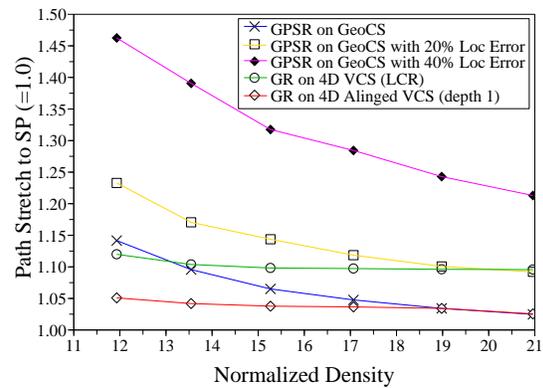}
\caption{Path Stretch of Reality: random deployment}
\label{fig:random_pl}
\end{figure}

The greedy ratio in these random scenarios is shown in Figure
\ref{fig:random_gr}. GeoCS performs very well even with 20\%
localization error. 4D VCS does not provide a higher greedy ratio than
GeoCS which may mean that in the reality, the geographical routing on
VCS may be not as good as GeoCS even with some localization
errors.  However, the greedy ratio with aligned VCS (even only with
depth 1) is higher than any other and approaches 100\% much quicker
than the others.  In our opinion, this result argues strongly in favor
of VCS, especially when localization error is common. In a relatively
sparse network, the high quality of aligned VCS makes it an attractive
solution for geographical routing.  Figure \ref{fig:random_pl} shows
the path length. Aligned VCS performs better than others except the
shortest path routing.

\section{Conclusion}
\label{sec:conclusion} 

In this paper, we first analyze the reason why the recently proposed
stateless routing protocols such as GPSR can not be successful as well
as traditional stateful routing protocols such as DSR and AODV since
the requirement of these routing protocols: precise location may not
be available mostly. We also analyze why the geographical routing on
VCS may not be a replacement of it on GeoCS since the performance of
it is not as well and the complementary solution for physical voids is
not satisfactory. Then we argue that the way of a stateless routing
protocol to be successful is to provide a performance as well as
stateful ones while does not suffer from the localization errors.

The simulation shows that the greedy forwarding phase of the stateless
geographical routing leads to a performance comparable with stateful
routing protocols.  We proposed an aligned virtual coordinate system
which significantly reduces the quantization noise in traditional VCS
and greatly increases the greedy ratio of these protocols.  AVCS was
shown to significantly outperforms GeoCS, basic VCS, and also the BVR
in terms of greedy ratio and path stretch.

It is widely accepted that stateless routing protocols offer desirable
properties over stateful ones.
However, it is also accepted that these come at a price in terms of
performance.  In this paper, we show that the performance of stateless
routing can approach that of stateful routing protocols, even
in the presence of voids.  Further, we show for the first time that
VCS (with alignment) can outperform GeoCS even without
localization errors.  


\begin{thebibliography}{1}
\bibitem{paper:aodv}
C. E. Perkins and E. M. Royer, \emph{Ad hoc On-Demand Distance
Vector Routing}, in Proceedings of the 2nd IEEE Workshop on Mobile
Computing Systems and Applications, February 1999

\bibitem{paper:gfg}
P. Bose, P. Morin, I. Stojmenovic and J. Urrutia,
\emph{Routing with guaranteed delivery in ad hoc wireless networks},
DIAL M99, Auguest 1999

\bibitem{paper:gpsr}
B. Karp and H.T. Kung,
\emph{{GPSR}: Greedy Perimeter Stateless Routing for Wireless Networks},
MobiCom 2000

\bibitem{paper:bphole}
Q. Fang, J. Gao and L. Guibas, \emph{Locating and Bypassing Routing
Holes in Sensor Networks}, INFOCOM 2004

\bibitem{paper:gcrp}
Sophia Fotopoulou-Prigipa and A. Bruce McDonald, \emph{GCRP:
Geographic Virtual Circuit Routing Protocol for Ad Hoc Networks},
MASS 2004

\bibitem{paper:practical_gr}
Young-Jin Kim, Ramesh Govindan, Brad Karp and Scott Shenker,
\emph{Geographic Routing Made Practical}, the Second USENIX/ACM
Symposium on Networked System Design and Implementation(NSDI'05),
May 2005


\bibitem{paper:errorgf}
T. He, C. Huang, B. Blum, J. A. Stankovic and T. Abdelzaher,
\emph{Range-Free Localization Schemes for Large Scale
  Sensor Networks}, MobiCom 2003

\bibitem{paper:errorfr}
K. Seada, A. Helmy and R. Govindan,
\emph{On the Effect of Localization Errors on Geographic Face Routing in
  Sensor Networks}, IPSN 2004



\bibitem{paper:grnoloc}
A. Rao, S. Ratnasamy, C. Papadimitriou, S. Shenker and Ion Stoica,
\emph{Geo. Routing without Location Info.}, MobiCom 2003

\bibitem{paper:gem}
J. Newsome and D. Song, \emph{GEM: Graph EMbedding for Routing and
Data-Centric Storage in
  Sensor Networks Without Geographic Information},
SenSys'03, Nov. 2003

\bibitem{paper:gfconjecture}
C. H. Papadimitriou and D. Ratajczak, \emph{On a Conjecture Related
to Geometric Routing}, ALGOSENSORS 2004

\bibitem{paper:vcembed}
T. Moscibroda, R. O'Dell, M. Wattenhofer, R. Wattenhofer,
\emph{Virtual Coordinates for Ad hoc and Sensor Networks},  ACM
DIALM-POMC,
October 2004

\bibitem{paper:lcr}
Qing Cao and Tarek F. Abdelzaher, \emph{A Scalable Logical
Coordinates Framework for Routing in Wireless Sensor Networks}, RTSS
2004

\bibitem{paper:vcsim}
D. M. Nicol, M. E. Goldsby and M. M. Johnson, \emph{Simulation
Analysis of Virtual Geographic Routing}, in Proceedings of the 2004
Winter Simulation Conference

\bibitem{paper:vcap}
A. Caruso et al.
\emph{GPS Free Coordinate Assignment and Routing in Wireless
Sensor Networks}, INFOCOM 2005

\bibitem{paper:bvr}
R. Fonseca, S. Ratnasamy, J. Zhao, C. T. Ee, D. Culler, S. Shenker
and I. Stoica, \emph{Beacon Vector Routing: Scalable Point-to-Point
Routing in Wireless Sensornets}, NSDI'05, 2005




\bibitem{paper:glider}
Q. Fang, J. Gao, L. J. Guibas, V. de Silva and L. Zhang,
\emph{GLIDER: Gradient Landmark-Based Distributed Routing for
  Sensor Networks}, INFOCOM 2005


\bibitem{paper:gc}
R. Nagpal, H. Shrobe and J. Bachrach,
\emph{Organizing a global coordinate system from local information
on an ad hoc sensor networks}, IPSN 2003

\bibitem{paper:aps}
D. Niculescu and B. Nath,
\emph{Ad Hoc Positioning System (APS)}, GlobalCom 2001


\bibitem{paper:hightower01}
J. Hightower and G. Borriella,
\emph{Location Systems for Ubiquitous Computing},
IEEE Computer, V34, 57-66, 2001

\bibitem{paper:convexposition01}
L. Doherty, K. S. J. Pister and L. El Ghaoui,
\emph{Convex position estimation in wireless sensor networks},
INFOCOM 2001

\bibitem{paper:robustloc-mobicom2002}
A. Haeberlen, E. Flannery, A. M. Ladd, A. Rudys, D. S. Wallach
and L. E. Kavraki,
\emph{Practical Robust Localization over Large-Scale {802.11} Wireless
  Networks}, MobiCom 2002

\bibitem{techrpt:hgr}
Ke Liu and Nael Abu-Ghazaleh, \emph{Virtual Coordinate Backtracking
for Void Traversal in Geographic Routing}, Technical report,
Binghamton University, CS. Dept., Feburary, 2006.


\bibitem{ns2}
\emph{The Network Simulator - ns-2}, http://www.isi.edu/nsnam/ns/







\end{thebibliography}
\end{document}